# Constraints on the nuclear symmetry energy and its density slope from the α decay process

W. M. Seif [(*)] and A. S. Hashem

*Cairo University, Faculty of Science, Department of Physics, 12613 Giza, Egypt*
[(*)] *wseif@sci.cu.edu.eg*

We study the impact of the nuclear symmetry energy and its density dependence on the α-decay process. Within the frame work of the performed cluster model and the energy density formalism, we use different parameterizations of the Skyrme energy density functionals that yield different equations of state EOSs. Each EOS is characterized by a particular symmetry-energy coefficient ($a_{sym}$) and a corresponding density-slope parameters $L$. The stepwise trends of the neutron (proton) skin thickness of the involved nuclei with both $a_{sym}$ and $L$ do not clarify the obtained oscillating behaviors of the α-decay half-life $T_\alpha$ with them. We find that the change of the skin thickness after α-decay satisfactory explains these behaviors. The presented results provide constrains on $a_{sym}$ centered around an optimum value $a_{sym}$= 32 MeV, and on $L$ between 41 and 57 MeV. These values of $a_{sym}$ and $L$, which indicate larger reduction of the proton-skin thickness and less increase in the neutron-skin thickness after an α-decay, yield minimum calculated half-life with the same extracted value of the α-preformation factor inside the parent nucleus.

## I. INTRODUCTION

To improve our knowledge of nuclear structure and nuclear properties of isospin-asymmetric nuclei and their interactions, we need to know accurate information about the symmetry energy and its density dependence. Various experimental and theoretical studies on nuclear structure, nuclear reactions, and nuclear astrophysics have investigated the symmetry energy and its density slope. For instance, $L$=58±18 MeV [1] has been obtained from comparing constraints from data on the neutron skin thickness of Sn isotopes, and those on isospin diffusion and double n/p ratio in heavy-ion collisions [1]. Analysis of other isospin diffusion data using an isospin- and momentum-dependent transport model, intermediate energy isoscaling measurements of heavy-ion collisions, and giant monopole resonance data have empirically indicated $L$=88±25 MeV [2]. The constraints $a_{sym}$ =30.5±3 MeV and $L$=52.5±20 MeV were determined from combining the symmetry energy at sub-saturation nuclear density and neutron skin thickness of Sn isotopes [3]. Simultaneous constraints on the baryon mass of a smaller mass member of a pulsar binary system and on modeling a progenitor star up to and through its collapse gave L≤ 70 MeV [4]. Consistent results from nuclear structure and heavy ion collisions data yielded a constraint centered around $a_{sym}$=32.5 MeV and $L$=70 MeV [5]. The values of $L$=66.5 MeV has been extracted from microscopic calculations based on realistic Argonne V18 NN potential and a phenomenological Urbana 3-body force [6]. $a_{sym}$=31.3 MeV and $L$=52.7 MeV were estimated using the Hugenholtz-Van Hove theorem and global nucleon optical potentials derived from single-particle energy levels, proton-neutron charge exchange reactions, and nucleon-nucleus scatterings [7]. The specified ranges of $a_{sym}$=30±5 MeV and 46≤$L$≤111 MeV were deduced using modified Skyrme-like model [8]. Based on the experimental pygmy strengths of Sn and Sb isotopes, the value of $a_{sym}$=32±1.8 was marked [9]. A value of $L$ = 64.8 ± 15.7 MeV has been provided from measurements of the pygmy dipole resonance on $^{132}$Sn and $^{68}$Ni [10]. Based on the dependence of recently measured neutron-removal cross sections of medium-heavy neutron-rich nuclei and their neutron skin, it was shown that $L$ could be constrained down to ±10 MeV [11]. However, the particularly indicated wide ranges of $L$ need more constraints from other investigations.

α-decays of heavy, super-heavy, and exotic nuclei had been used in different studies to explore diverse nuclear structure and related quantities. For instance, the α-decay process has been used to probe the nuclear incompressibility [12], the neutron and proton shell closures of finite nuclei [13,14,15], and the spin-parity configuration [16,17,18] of their ground- and isomeric states. Also, it has been used to investigate the collective vibrational and rotational excitations [19] of nuclei and their deformations [20,21]. It was found that considering the neutron-skin thickness of the daughter nucleus reduces the calculated half-life against α [22] and cluster [23] decays. On the other hand, it was recently concluded that the proton-skin thickness ($\Delta_p$) also reduces the stability of the nucleus and decreases its half-life against α-decay [24]. Along the same isotopic chain, it was found that the half-lives of the proton-skinned isotopes exponentially decrease with increasing $\Delta_p$, while the $Q_\alpha$-value linearly increases with it [24]. Attempts were made to constrain the quadratic and quartic symmetry energies, their density slopes, and the neutron skin thickness of $^{208}$Pb via cluster radioactivity [25,26]. Recently, a study has been shown that both the half-life of an α-decay and its released energy consistently follow the change of proton (neutron) skins, from parent to daughter nuclei [27]. It was indicated that the α-decays of the proton- (neutron-) skinned nuclei typically proceed to produce a significant decrease (a very least increase) in the thickness of the proton (neutron) skins of daughter nuclei. As the proton- (neutron-) skin thickness of the nucleus directly correlates with the nuclear symmetry energy and its density-slope, we try in the present work to get more strict restriction on the symmetry energy and its density-slope from the α-decay process. In the following section, we outline the general formalism for calculating the α-decay penetration probability and half-life, based on different nuclear equations of state in the framework of the Wentzel-Kramers-Brillouin approximation and the preformed cluster model. In Sec. III we present and analyze our results for the α-decay process of the $^{105}$Te and $^{212}$Po nuclei. Finally, a brief summary and our main conclusions are given in Sec. IV.

## II. THEORETICAL FORMALISM

In the Skyrme energy density formalism, the total energy density functional (EDF) reads [28,29]

$$H(\rho_i,\tau_i,J_i) = \sum_{i=n,p} \frac{\hbar^2}{2m_i}\tau_i\,(\rho_i,\nabla\rho_i,\nabla^2\rho_i) + H_{Sky}(\rho_i,\tau_i,J_i) + H_{Coul}(\rho_p). \qquad (1)$$

While the first term on the right-hand side of Eq. (1) corresponds to the kinetic energy, the second term defines the Skyrme nuclear energy. They are given in terms of the proton $\rho_p$ and neutron $\rho_n$ densities, and the corresponding kinetic energy $\tau_i$ ($i$=$p,n$) and spin-orbit $J_i$ densities. Both $\tau_i$ and $J_i$ can be calculated using the extended Thomas-Fermi approximation [30] as functions of $\rho_i$, $\nabla\rho_i$, $\nabla^2\rho_i$, and $f_i(\vec{r}) = m_i/m_i^{eff}(\vec{r})$. While $m_{i=\mathrm{p,n}}$ are the proton and neutron masses, $m_i^{eff}$ represent their effective mass. $H_{coul}$ is the Coulomb energy density. The nuclear and Coulomb parts, respectively, of the EDF can take the explicit forms [31,32],

$$H_{Sky} = \frac{t_0}{2}\left[\left(\frac{x_0}{2}+1\right)\rho^2 - \left(x_0+\frac{1}{2}\right)\sum_{i=p,n}\rho_i^2\right] + \frac{t_3\rho^\sigma}{12}\left[\left(\frac{x_3}{2}+1\right)\rho^2 - \left(x_3+\frac{1}{2}\right)\sum_{i=p,n}\rho_i^2\right]$$
$$+\frac{1}{4}\left[t_2\left(\frac{x_2}{2}+1\right)+t_1\left(\frac{x_1}{2}+1\right)\right]\rho\tau + \frac{1}{4}\left[t_2\left(x_2+\frac{1}{2}\right)-t_1\left(x_1+\frac{1}{2}\right)\right]\sum_{i=p,n}\rho_i\tau_i$$
$$-\frac{1}{16}\left[t_2\left(\frac{x_2}{2}+1\right)-3t_1\left(\frac{x_1}{2}+1\right)\right](\nabla\rho)^2$$
$$-\frac{1}{16}\left[t_2\left(x_2+\frac{1}{2}\right)+3t_1\left(x_1+\frac{1}{2}\right)\right]\sum_{i=p,n}(\nabla\rho_i)^2$$
$$+\frac{1}{16}\left[(t_1-t_2)\sum_{i=p,n}J_i^2-(t_1x_1+t_2x_2)J^2\right]+\frac{W_0}{2}\left(J\cdot\nabla\rho+\sum_{i=p,n}J_i\cdot\nabla\rho_i\right), \quad (2)$$

and

$$H_{Coul} = H_C^{dir} + H_C^{exch} = \frac{e^2}{2}\rho_p(\vec{r})\int\frac{\rho_p(\vec{r}')}{|\vec{r}-\vec{r}'|}d\vec{r}' - \frac{3e^2}{4}\left(\frac{3}{\pi}\right)^{\frac{1}{3}}\left(\rho_p(\vec{r})\right)^{\frac{4}{3}}. \quad (3)$$

Here $t_i(i=0,1,2,3)$, $x_i$, σ, and $W_0$ represent the Skyrme force parameters. $H_C^{dir}$ and $H_C^{exch}$ define the direct and exchange parts, respectively, of the Coulomb EDF.

Based on the EDF given by Eqs. (1)-(3) and the frozen density approximation, we can obtain the interaction potential between the emitted α-particle and the daughter nucleus as a function of the separation distance $r$ between their centers of mass as [15,29,33,34],

$$V(r) = \int\left\{H\left[\rho_{p\alpha}(\vec{x})+\rho_{pD}(r,\vec{x}),\rho_{n\alpha}(\vec{x})+\rho_{nD}(r,\vec{x})\right]-H_\alpha\left[\rho_{p\alpha}(\vec{x}),\rho_{n\alpha}(\vec{x})\right]\right.$$
$$\left.-H_D\left[\rho_{pD}(\vec{x}),\rho_{nD}(\vec{x})\right]\right\}d\vec{x}\,. \quad (4)$$

$H$, $H_\alpha$ and $H_D$ define the EDF of the composite system and that of the individual α and daughter (D) nuclei, respectively. $\rho_{ij}(i=p,n;j=\alpha,D)$ represent the corresponding protons and neutrons density distributions. These density distributions will be self-consistently determined by Hartree-Fock calculations [35,36], based on the different considered parameterizations of EDF. The multipole expansion method [37,38] will be used to compute the direct part of the Coulomb potential, which involves the finite range p-p Coulomb interaction, Eq. (3). More details concerning the method of calculations can be found in Refs. [17,33]. From the self-consistently determined proton and neutron density distributions of a given nucleus, one can estimate its neutron- (proton-) skin thickness as,

$$\Delta_{n(p)}(A,Z) = R_{n(p)}^{rms}(A,Z) - R_{p(n)}^{rms}(A,Z), \quad (5)$$

where the neutron (proton) rms radius reads

$$R_{n(p)}^{rms} = \langle R_{n(p)}^2 \rangle^{1/2} = \left( \frac{\int r_{n(p)}^2 \rho_{n(p)}(\vec{r}) d\vec{r}}{\int \rho_{n(p)}(\vec{r}) d\vec{r}} \right)^{1/2}.$$

Considering an infinite asymmetric nuclear matter (ANM), we can write the energy per nucleon of ANM with a the proton fraction $\eta=Z/A$ in terms of the Skyrme EDF as [32],

$$\begin{aligned} E_A = \frac{H(\rho)}{\rho} = \frac{3\hbar^2}{10m} k_F^2 H_{5/3} + \frac{t_0}{4}\rho\left[(x_0+2) - \left(x_0 + \frac{1}{2}\right) H_2\right] \\ + \frac{t_3 \rho^{\sigma+1}}{24}\left[(x_3+2) - \left(x_3+\frac{1}{2}\right)H_2\right] \\ + \frac{3k_F^2}{40}\left\{(2t_1+2t_2+t_1x_1+t_2x_2)\rho H_{5/3} + \left(\frac{t_2}{2} - \frac{t_1}{2} + t_2x_2 - t_1x_1\right)\rho H_{8/3}\right\}, \quad (6) \end{aligned}$$

where $k_F=(3\pi^2\rho/2)^{1/3}$ and $H_n(\eta) = 2^{n-1}[\eta^n + (1-\eta)^n]$. Expanding the equation of state given by Eq. (6) as a function of $\eta$ and $\rho$, we can define the symmetry energy $E_{sym}$ that measures the isospin dependence of the nucleon-nucleon (NN) interaction as,

$$\begin{aligned} E_{sym}(\rho) = \frac{1}{8}\frac{\partial^2 E_A(\rho,\eta)}{\partial \eta^2}\bigg|_{\eta=\frac{1}{2}} \\ = \frac{\hbar^2 k_F^2}{6m} - \frac{t_0}{4}\left(x_0+\frac{1}{2}\right)\rho - \frac{t_3}{24}\left(x_3+\frac{1}{2}\right)\rho^{\sigma+1} + \frac{k_F^2}{24}\{(4t_2 - 3t_1x_1 + 5t_2x_2)\rho\}. \quad (7) \end{aligned}$$

One of the characteristic quantities for the EOS is the symmetry energy coefficient $a_{sym}=E_{sym}(\rho_0)$ that defined at normal saturation density $\rho_0$. Another important quantity associated with the symmetry energy is the slope $L$ of its density dependence. It can be written in the form

$$\begin{aligned} L = 3\rho_0 \frac{\partial E_{sym}(\rho)}{\partial \rho}\bigg|_{\rho_0} \\ = \frac{\hbar^2 k_{F0}^2}{3m} - \frac{3t_0}{4}\left(x_0+\frac{1}{2}\right)\rho_0 - \frac{t_3}{8}\left(x_3+\frac{1}{2}\right)(\sigma+1){\rho_0}^{\sigma+1} \\ + \frac{k_{F0}^2}{24}[5(4t_2 - 3t_1x_1 + 5t_2x_2)\rho_0], \quad (8) \end{aligned}$$

with $k_{F0}=(3\pi^2\rho_0/2)^{1/3}$.

In the performed cluster model [39,40], the α decay-width is given in terms of the assault frequency $\nu$ and the penetration probability $P$ of the tunneling process as

$$\Gamma = \hbar \nu P. \quad (9)$$

We can find the assault frequency and the penetration probability using the Wentzel-Kramers-Brillouin approximation, respectively, as

$$\nu = T^{-1} = \left[ \int_{R_1}^{R_2} \frac{2\mu}{\hbar k(r)} dr \right]^{-1}, \tag{10}$$

and

$$P = \exp\left( -2 \int_{R_2}^{R_3} k(r) dr \right), \tag{11}$$

where $k(r) = \sqrt{2\mu |V(r) - Q_\alpha| / \hbar^2}$. $Q_\alpha$ is the $Q$-value of the decay process. The experimental values of $Q_\alpha$ will be used in the present calculations. $\mu = m_\alpha\ m_D\ /\ (m_\alpha + m_D)$ defines the reduced mass of the $\alpha$ ($m_\alpha$) and daughter ($m_D$) system. The three classical turning points $R_{i=1,2,3}$ (fm) are defined along the path of emitted $\alpha$-particle with respect to the daughter nucleus as $V(r)|_{r=R_i} = Q_\alpha$. For the unfavored decays between different spin-parity assignments of the patent and daughter nuclei, a centrifugal part is added to the total potential given by Eq. (4), to take into account the angular momentum transferred by the emitted $\alpha$-particle. The decays considered in the present work are favored decays with no transferred angular momentum. Now, we can estimate the half-life against α-decay in terms of the calculated decay width and the preformation factor $S_\alpha$ of the α-particle in the parent nucleus as,

$$T_\alpha = \frac{\hbar\, ln\, 2}{S_\alpha \Gamma}. \tag{12}$$

The preformation factor $S_\alpha$ can be obtained microscopically [41,42], semi-microscopically [43], or using some available semi-empirical formulas [17,18].

## III. RESULTS AND DISCUSSION:

In this section, we investigate the effects of the nuclear symmetry energy and its density dependence on the α-decay process of both $^{105}$Te and $^{212}$Po nuclei. To do so we used many Skyrme NN interactions yielding different equations of state. While the $^{105}$Te nucleus and its $^{101}$Sn daughter nucleus have proton-skin thickness, both $^{212}$Po and $^{208}$Pb possesses neutron-skin thickness. Both the $^{101}$Sn ($Z$=50) and $^{208}$Pb ($Z$=82) daughter nuclei possess proton closed proton shell and their density distributions are almost spherical. $^{208}$Pb has a closed neutron shell as well, $N$=126.

Fig. 1 shows the influence of the nuclear symmetry energy on the calculated α-decay half-life $T_\alpha$. The calculations displayed in Fig. 1 are carried out using twenty three parameterizations of the Skyrme EDFs. Namely, the SkSc10 [44], SkSc6 [45], SkSC1-3 [45], SkM1 [45], Es [47], RATP [48], SkSc14 [49], SkSc5 [44], SkT3 [50], SLy4 [31], KDEX [51], KDE0v [52], SkI2 [53], SII [28], KDE0v1 [52], Skxs20 [54], SkI5 [53], Ska35s25 [55], SK272 [56], Skxs25 [54], and SGOI [57] parameterizations have been used. These parameterizations yield equations of state characterized by a symmetry energy coefficient ranges from $a_{sym}$(SkSc10) = 22.83 MeV to $a_{sym}$(SGOI) = 45.20 MeV. Plotted in Figs. 1(a) and 1(b), respectively, are the calculated half-lives of the $^{105}$Te and the $^{212}$Po nuclei, without introducing the preformation factor $S_\alpha$, as functions of $a_{sym}$. As seen in Fig. 1, there is a hesitant one-to-one correspondence between the calculated half-life and $a_{sym}$. This is expected because of the influences of the other EOS properties such as the incompressibility and the surface-energy coefficients. The calculations based on the Skyrme-SLy4 ($a_{sym}$=32 MeV) force indicate the minimal calculated half-lives for both displayed cases. Any deviation from this value, either by increasing or by decreasing $a_{sym}$, yields larger half-life.

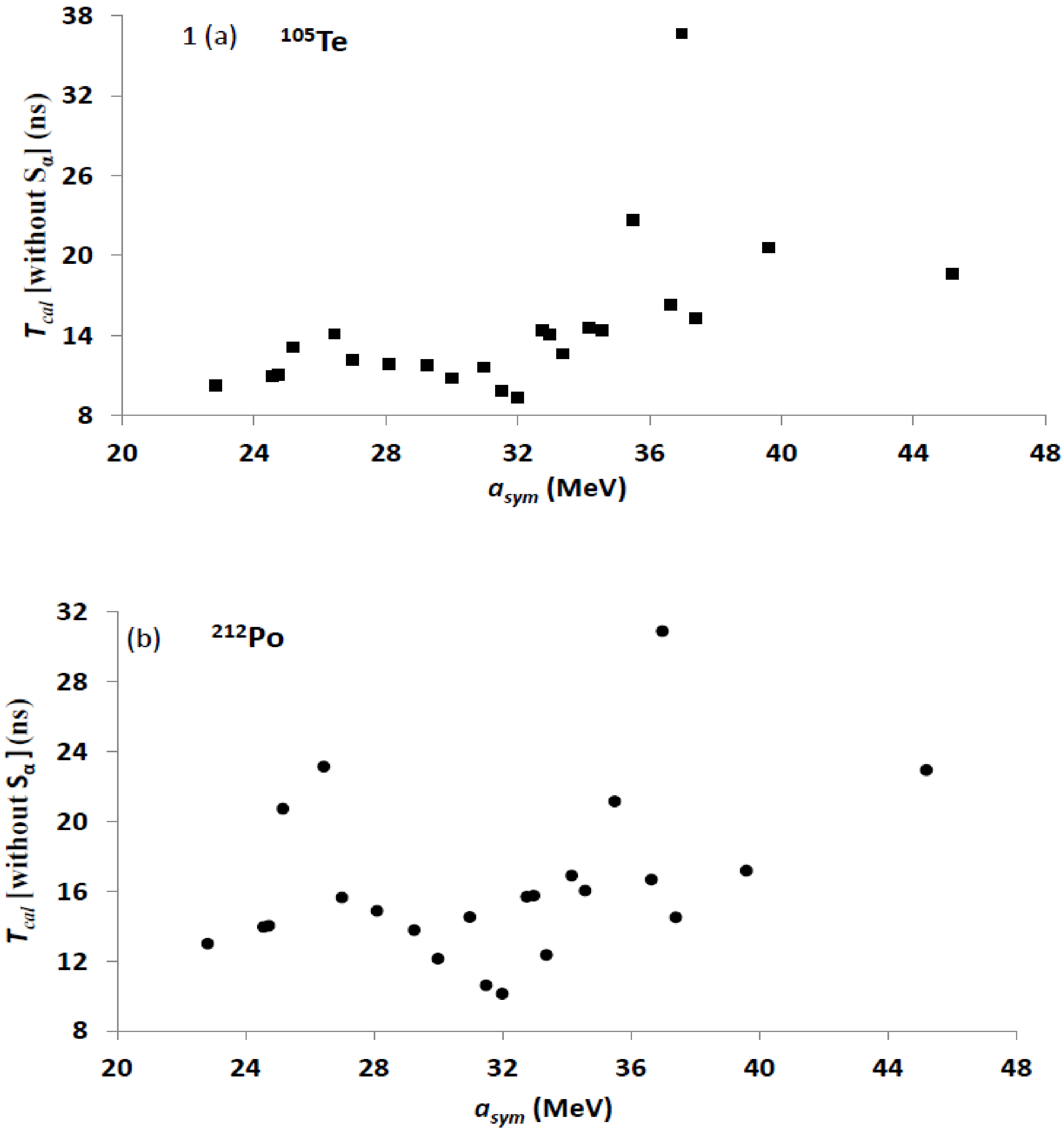

**Fig. 1**: The calculated partial half-life $T_\alpha$ (Eq. (12) without introducing the preformation factor $S_\alpha$) of the ground-state to ground-state α decay of (a) $^{105}$Te and (b) $^{212}$Po nuclei, as a function of the symmetry energy coefficient $a_{\rm sym}$ corresponding to the used EDF.

The characteristics of the symmetry energy of a given EOS are determined not only by the symmetry energy coefficient that measures the isovector curvature of the EOS at saturation density, but also by the slope ($L$) of the symmetry energy as a function of density. We thus need to check the effect of the density-slope of the symmetry energy on the calculated half-life. To achieve this, we have used different Skyrme interactions that generate equations of state of the same symmetry energy coefficient but with different corresponding density-slopes. Figures 2(a) and 2(b) show the calculated half-lives of $^{105}$Te and $^{212}$Po, respectively, as functions of the density-slope of the symmetry energy $L$(MeV). Twenty nine EDFs have been used to perform the calculations presented in Fig. 2, namely the Skz1-4 [58], Skxs15 [54], SLy230a [59], SLy1-3 [60], SLy4-5 [31], SLy9 [60], SLy10 [31], KDE [52], FPLyon [61], T12 [62], T32 [62], T63 [62], SkT2-3 [50], SV-sym32 [63], NRAPR [64], Ska25s20 [55], Ska35s20 [55], Ska45s20 [55], SkO' [65], SkA [66], Sefm074 [67], and GS [47] EDFs. These EDFs generate EOSs characterized by a narrow range of symmetry energy coefficient $a_{sym}$=32.4±1.4 MeV, but with a rather wide range of $L$ from $L$(Skz4) = 5.75 MeV to $L$(GS) = 93.31 MeV. Figure 2 shows that the calculated half-life fluctuates over the different regions of the density-slope of the symmetry energy, without changing it order of magnitude. However, the range of $L$= 41 MeV - 57 MeV averagely yields the same calculated half-life, $T_\alpha$(without $S_\alpha$) = 9±1 ns and 10±2 ns for $^{105}$Te and $^{212}$Po, respectively. $T_\alpha$ considerably increases in the neighborhood before and after this range of $L$, then it starts to decrease again.

As mentioned above, the proton- and neutron-skin thicknesses of nuclei strongly correlate with the symmetry energy and its density-slope [68,69]. Meanwhile, the α-decay half-lives directly correlate with the change of the proton (neutron) skin thickness after decays [27]. Now, the question arises whether the behavior of the calculated half-life presented in Figs. 1 and 2 is due to the effect of the proton (neutron) skin thickness of the participating nuclei, or to the change of skin thickness after α-decay. To answer this question, we plot in Fig. 3(a) the proton $\Delta_p$ (neutron $\Delta_n$) skin thickness of the $^{105}$Te ($^{212}$Po) parent nucleus and that of the corresponding $^{101}$Sn ($^{208}$Pb) daughter nucleus, as functions of the symmetry energy coefficient. The presented $\Delta_{p(n)}$ are self-consistently calculated using HFB method [35,36], based on the Skyrme EDFs used in Fig. 1. Figure 3(a) shows that the proton-skin thickness of both $^{105}$Te and $^{101}$Sn slightly decreases as $a_{sym}$ increases. The $^{101}$Sn daughter nucleus has less proton-skin thickness than that of the $^{105}$Te parent nucleus. On the other hand, the neutron-skin thickness of both $^{212}$Po and $^{208}$Pb sharply increases with $a_{sym}$. The $^{208}$Pb daughter nucleus possesses larger neutron-skin thickness relative to that of the $^{208}$Pb parent nucleus. Clearly, the steady behavior of $\Delta_{p(n)}$ with $a_{sym}$ could not be a reason for the oscillating behaviors shown in Figs. 1(a) and 1(b). Shown in Figs. 3(b) and 3(c) are, respectively, the decrease of the proton-skin thickness and the increase of the neutron-skin thickness after the α-decays of $^{105}$Te and $^{212}$Po, $\delta_{i\alpha}(i\text{=p,n})=\Delta_i$(daughter nucleus)$-\Delta_i$ (parent nucleus), as functions of $a_{sym}$. Figures 3(b) and 3(c) show fluctuating behaviors of $\delta_{p\alpha}(^{101}\text{Sn},^{105}\text{Te})$ and $\delta_{n\alpha}(^{208}\text{Pb}, ^{212}\text{Po})$ with the symmetry energy coefficient, similar to that of $T_\alpha$ with $a_{sym}$ in Fig. 1. The maximal decrease in the proton-skin and the minimal increase in the neutron-skin after the α-decays of $^{105}$Te and $^{212}$Po, respectively, are both obtained at $a_{sym}$=32 MeV, which yielded the minimal $T_\alpha$ in Figs. 1(a) and 1(b). This is consistent with the conclusions of Ref. [27], which indicated that the α–decays of the proton (neutron) skinned parent nuclei preferably proceed to yield a significant reduction (very least increase) in the proton (neutron) skin thickness of their daughter nuclei. So, the values of $a_{sym}$ =32 MeV as indicated in Fig. 1 and in Figs. 3(b) and 3(c) can be marked as the central optimum value of the symmetry energy coefficient towards producing more stable nucleus in an α-decay process. This marked value of $a_{sym}$ lies at the center of the range indicated from the measured pygmy strengths of Sn and Sb isotopes ($a_{sym}$=32±1.8 [9]). It is also included within the range extracted from investigating the neutron skin thickness of Sn isotopes ($a_{sym}$=30.5±3 MeV [3]) and that obtained using modified Skyrme-like model ($a_{sym}$=30±5 MeV [8]). Moreover, it is consistent with the results extracted from nuclear structure and heavy ion collisions analysis, which is around $a_{sym}$=32.5 MeV [5], and the value indicated by derived optical potentials ($a_{sym}$=31.3 MeV [7]).

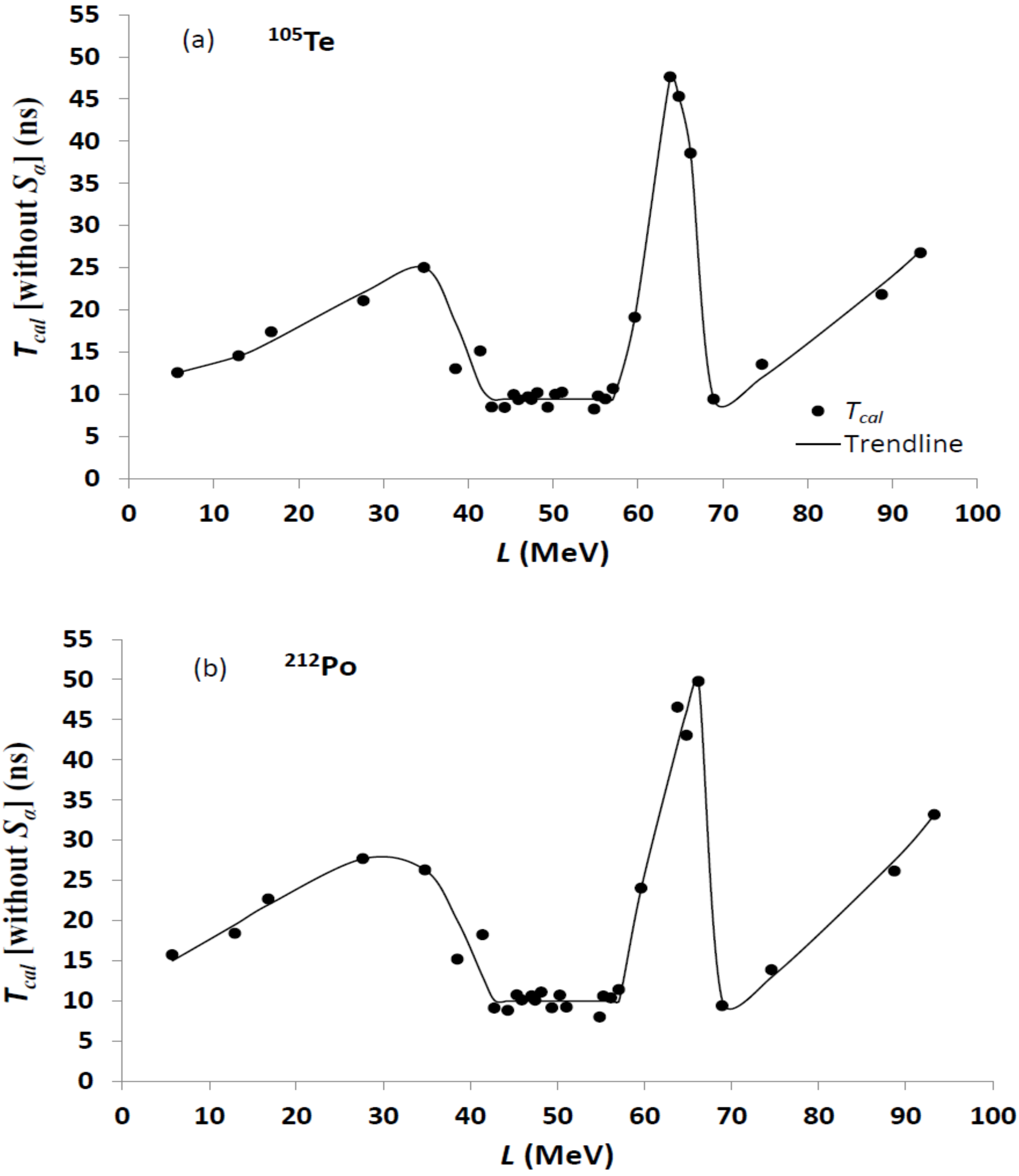

**Fig. 2**: The same as Fig. 1 but the calculated $T_\alpha$ is displayed as a function of the density slope $L$ of the nuclear symmetry energy.

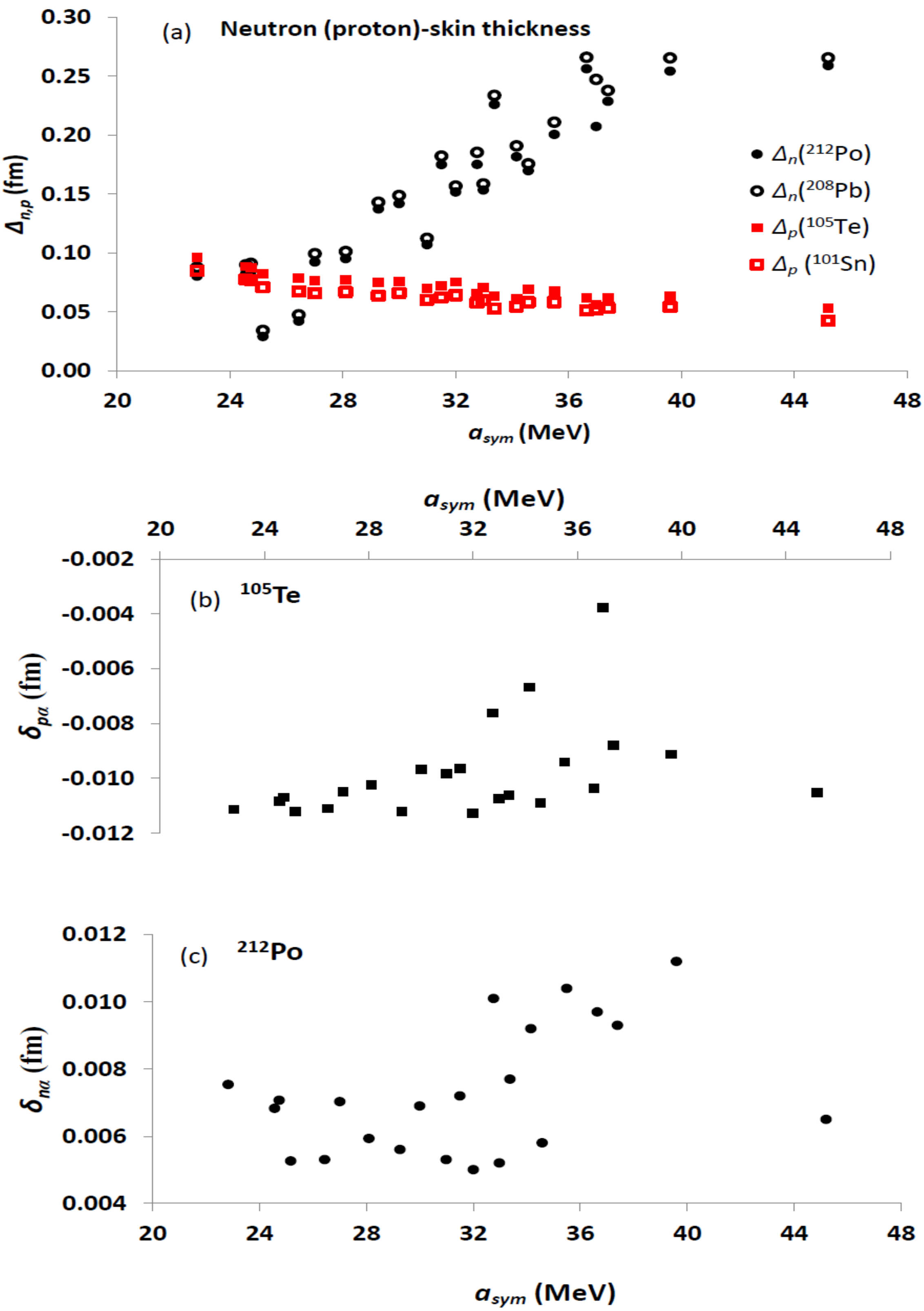

**Fig. 3**: (a) The proton $\Delta_p$ (neutron $\Delta_n$) skin thickness of the $^{105}$Te ($^{212}$Po) parent nucleus and that of its $^{101}$Sn ($^{208}$Pb) daughter nucleus, as functions of the symmetry energy coefficient. Panels (b) and (c) show the decrease of $\Delta_p$ after the α-decay of $^{105}$Te and the increase of $\Delta_n$ after the α-decay of $^{212}$Po. The displayed quantities are calculated based on the EDFs used in Fig. 1.

Fig. 4(a) shows the proton- and neutron-skin thicknesses of the $^{105}$Te and $^{212}$Po α-emitters, respectively, and of their daughter nuclei $^{101}$Sn and $^{208}$Pb, as functions of the density-slope of the symmetry energy. The $\Delta_{p(n)}$ displayed in Fig. 4 are calculated in terms of the Skyrme EDFs used in Fig. 2. As shown in Fig. 4(a), while the proton-skin thicknesses of $^{105}$Te and $^{101}$Sn are almost independent of $L$, the neutron-skin thicknesses of $^{212}$Po and $^{208}$Pb show increasing behavior with $L$. Again, the oscillating behavior of $T_\alpha$ with $L$ as shown in Figs. 2(a) and 2(b) cannot be explained by the steadily behavior of $\Delta_{p(n)}$ with $L$. Displayed in Figs. 4(b) and 4(c) are the decrease in $\Delta_p$ and the increase in $\Delta_n$ after the α-decays of $^{105}$Te and $^{212}$Po, respectively, as functions of $L$. Comparing Fig. 2 and Fig. 4 one can observe that the range $L$= 41-57 MeV that yields almost constant minimum values of $T_\alpha$ in Figs. 2(a) and 2(b) exhibits the larger decrease in $\Delta_p$ (Figs. 4(a)) and the smaller increase in $\Delta_n$ (Figs. 4(b)) after the α-decays of $^{105}$Te and $^{212}$Po, respectively. This indicated range of $L$ completely overlaps with the constrained ranges that extracted from the isospin diffusion data ($L$=58±18 MeV [1]), from the radioactivity of proton emitters ($L$=51.8±7.2 MeV [70]), from the neutron skin thickness of Sn isotopes ($L$=52.5±20 MeV [3]), and from the pygmy dipole resonance of $^{68}$Ni and $^{132}$Sn ($L$ = 64.8 ± 15.7 MeV) [10]. The range of $L$ indicated in Figs. 2 and 4 is also consistent with neutron stars investigations ($L \leq$ 70 MeV [4]), with analysis of optical potentials extracted from nuclear structure and reactions ($L$=52.7 MeV [7]), and with the results based on modified Skyrme-like model (46≤$L$≤111 MeV [8]). The range of $L$ provided here remarkably limits these indicated wide ranges.

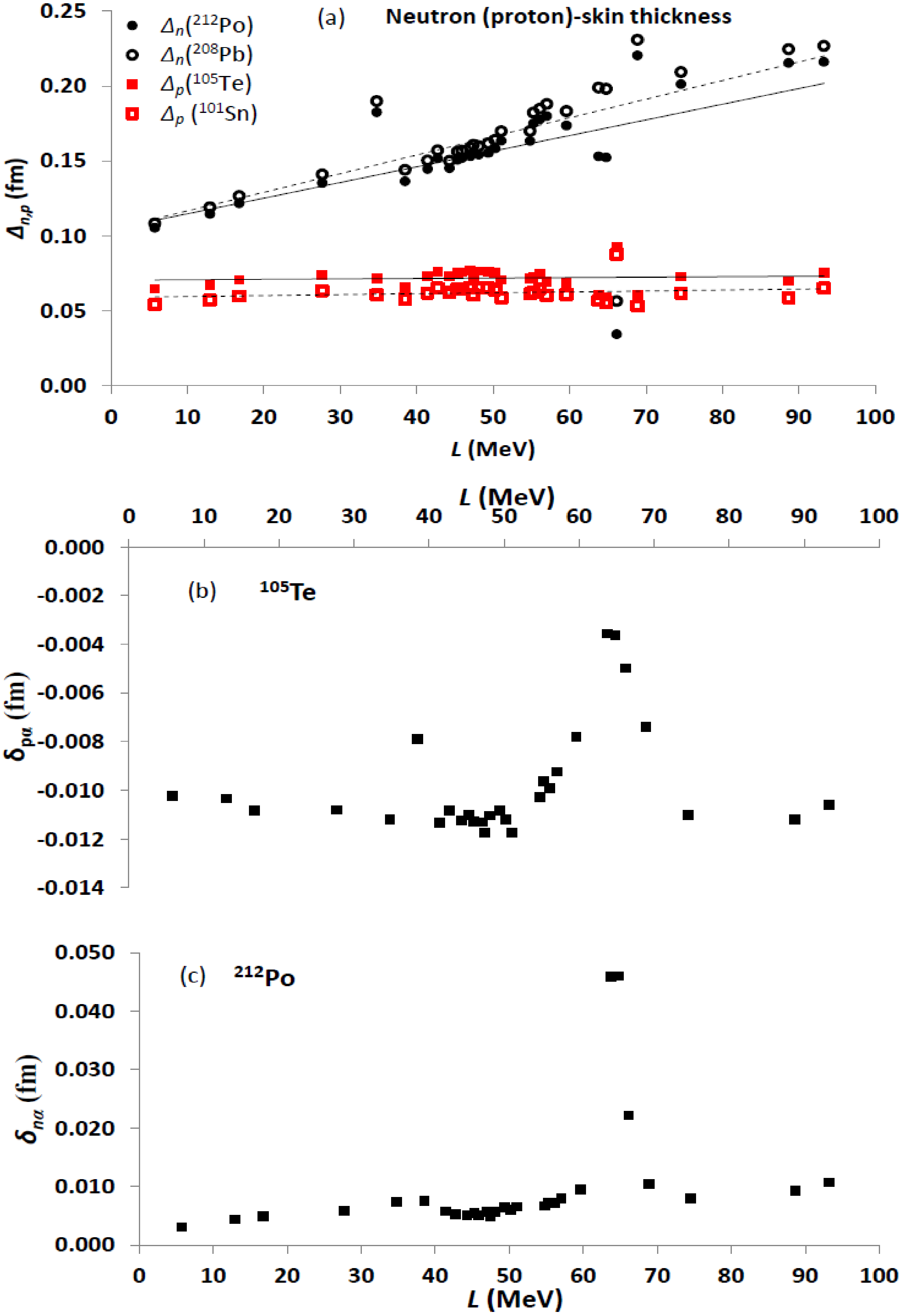

**Fig. 4**: The same as Fig. 3 but the shown quantities are displayed as functions of the density slope *L* of the symmetry energy. The displayed quantities are calculated based on the EDFs used in Fig. 2.

Figure 5 shows the effect of the symmetry energy on the α-core interaction potential at the different separation distances $r$(fm) between the centers of mass of the interacting nuclei. Figures 5(a) and 5(b) show the nuclear and total potentials, respectively, between an α particle and $^{208}$Pb daughter nucleus, which are involved in the α-decay of $^{212}$Po ($Q_\alpha$=8.954 MeV [71]). The calculations presented in Fig. 5 are performed using the Es ($a_{sym}$=26.44 MeV), SLy4 (32.0 MeV) and Ska35s25 (36.98 MeV) parameterizations of the Skyrme EDFs. Figure 5(a) shows that increasing the symmetry energy increases the attractive nuclear part in the fully-overlapped density region of the interaction potential, at which $r$ is less than the sum of the radii of the two interacting nuclei. The effect of the symmetry energy decreases in the surface and tail regions of the nuclear potential. The change of the symmetry energy slightly affects the repulsive Coulomb potential. As a result, both the width and depth of the internal pocket of the total potential increase with increasing the symmetry energy coefficient, as seen in Fig. 5(b). This seriously affects the preformation probability of an α particle near the surface of the parent nucleus and decreases its assault frequency, Eq.(10). The competition between the symmetry and Coulomb energies weakens the effect of the change in the symmetry energy near the Coulomb barrier. However, the shift in the position of the second turning point $R_2$ that located around the surface region of the interaction potential with the change of the symmetry energy affects the penetration probability, Eq. (11). The balance between the symmetry and the Coulomb energy yield the optimum value of symmetry energy coefficient towards more stability.

Finally, we show in Figs. 6(a) and 6(b) the preformation factor $S_\alpha$ of the α-particle inside the $^{105}$Te and $^{212}$Po nuclei, respectively, as extracted from their experimental half-lives and their calculated half-lives without introducing $S_\alpha$, Eq. (12). The estimated values of $S_\alpha$ are displayed as functions of the symmetry energy slope parameter $L$ that related to the used Skyrme interaction. The calculation presented in Fig. 6 were performed using the EDFs that have been used in Fig. 2 but yield narrower investigated range of $L$, from $L$ (Skz1)=27.67 MeV to $L$(Ska45s20)=66.21 MeV. The uncertainties in the $Q_\alpha$-value [71] and in the experimental half-life [72] are both taken into account in the extracted values of $S_\alpha$. As seen in Fig. 6, the range of $L$= 41-57 MeV yields an average constant values of $S_\alpha(^{105}\text{Te}) = 0.016\pm0.003$ and $S_\alpha(^{212}\text{Po}) = 0.033\pm0.007$. We recall here that the estimation of the preformation factor is model dependent [15]. For instance, several values of $S_\alpha(^{212}\text{Po})$ have been extracted based on different models [41,73,74,75]. So, our indicated constrains on $a_{sym}$ and $L$ rely on the obtained maximum reduction (less increase) of the proton- (neutron-) skin thickness after an α-decay, and the obtained minimum calculated $T_\alpha$ with the same extracted value of the α-preformation factor, but not on the calculated values themselves of $T_\alpha$ and $S_\alpha$.

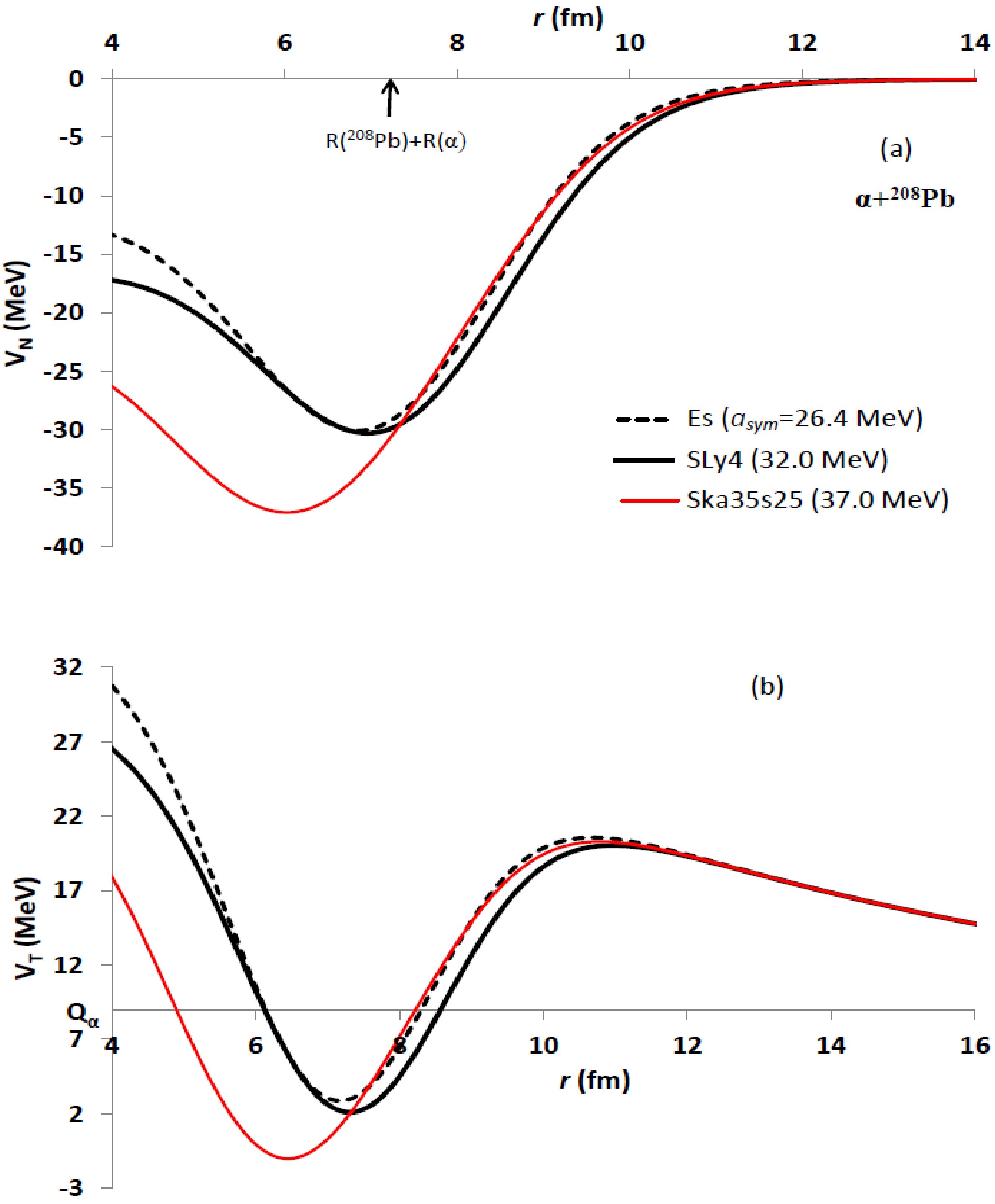

**Fig. 5**: The radial dependence of the (a) nuclear and (b) total interaction potential between α and $^{208}$Pb nuclei, which are participating in the α decay of $^{212}$Po, based on three Skyrme EDFs yielding different values of the symmetry energy coefficient.

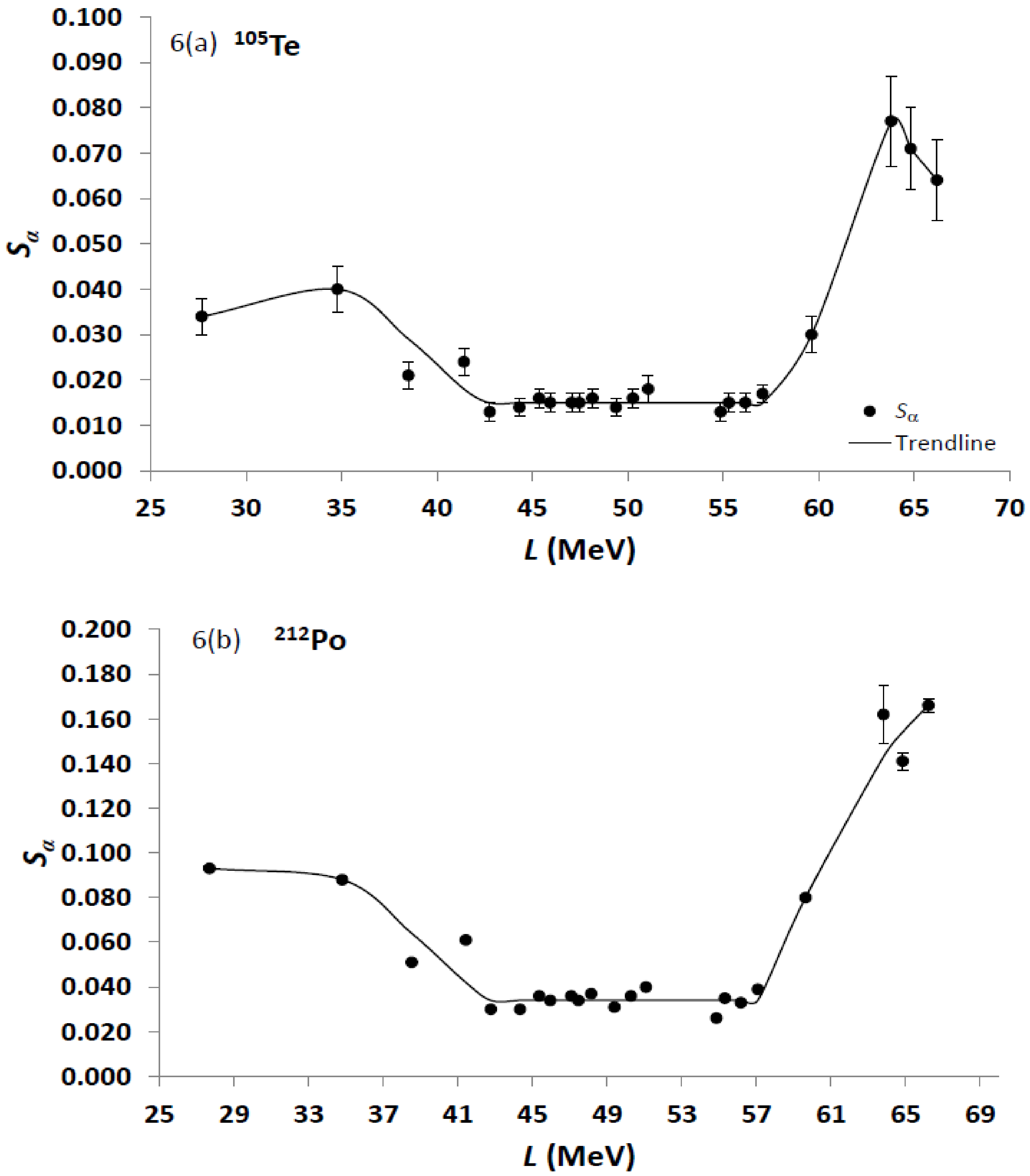

**Fig. 6**: The α-preformation factor inside (a) $^{105}$Te and (b) $^{212}$Po α-emitters, as a function of the density slope of the symmetry energy. The values of $S_\alpha$ are extracted from the observed half-lives and the calculated half-lives without introducing $S_\alpha$ (Eq. (12)). The uncertainties in both the $Q_\alpha$-value and the observed half-life are considered in the extracted value of $S_\alpha$.

## IV. SUMMARY AND CONCLUSIONS

In this work, we have studied the impact of the nuclear symmetry energy and its density dependence on the α decays of the $^{105}$Te and $^{212}$Po nuclei. We have used a total of 50 Skyrme EDFs yielding different equations of state characterized by a symmetry energy coefficient $a_{sym}$= 22.83 - 45.20 MeV, and corresponding density-slope ranges from $L$= -36.86 MeV to 129.3 MeV. We have found that the symmetry energy increases the attractive nuclear part of the total potential in the fully-overlapped density region. This increases both the width and depth of the internal pocket of the total potential, which in turn affects the α preformation probability and decreases its assault frequency. The balance between the symmetry and the Coulomb energies weakens its effect near the Coulomb barrier and yields the optimum value of symmetry energy coefficient towards more stability.

The calculations based on the Skyrme EDF characterized by $a_{sym}$=32 MeV have yielded the minimal calculated half-lives of both $^{105}$Te and $^{212}$Po. The values of $L$ within the range 41 MeV≤ $L \leq$ 57 MeV have averagely yielded the same calculated half-life. $T_{\alpha}$ considerably increased in the neighborhood outside this range of $L$, then it began to decrease again. The proton-skin thickness has shown slightly decreasing behavior with $a_{sym}$ and almost independence of $L$. The neutron-skin thickness has shown increasing trends with both $a_{sym}$ and $L$. These stepwise trends of $\Delta_{p(n)}$ with both $a_{sym}$ and $L$ did not explain the oscillating behaviors of $T_{\alpha}$ with them. Meanwhile, the change of the proton or neutron skin thickness from the parent to daughter nuclei has shown fluctuating behavior with $a_{sym}$. The maximal reduction of the proton-skin thickness and the minimal rise in the neutron-skin thickness after the α-decays of $^{105}$Te and $^{212}$Po, respectively, have been obtained at $a_{sym}$=32 MeV, which has indicated the minimal $T_{\alpha}$. Also, the range of $L$ between 41 and 57 MeV, which yielded the least calculated values of $T_{\alpha}$, have exhibited the larger reduction in $\Delta_p$ and the smaller increase in $\Delta_n$ after the α-decay. This range of $L$ have yielded an average constant value of α-preformation factor in the parent nucleus, $S_{\alpha}(^{105}\mathrm{Te})$ = 0.016±0.003 and $S_{\alpha}(^{212}\mathrm{Po})$ = 0.033±0.007.

## REFERENCES

[1] Lie-Wen Chen, Che Ming Ko, Bao-An Li, and Jun Xu, Phys. Rev. C **82**, 024321 (2010).

[2] L.W. Chen, C.M. Ko, and B.A. Li, Phys. Rev. C **76**, 054316 (2007).

[3] Lie-Wen Chen, Phys. Rev. C **83**, 044308 (2011).

[4] W. G. Newton and B. A. Li, Phys. Rev. C **80**, 065809 (2009).

[5] M. B. Tsang *et al*., Phys. Rev. C **86**, 015803 (2012).

[6] I. Vidana, C. Providencia, A. Polls, and A. Rios, Phys. Rev. C **80**, 045806 (2009).

[7] Chang Xu, Bao-An Li, and Lie-Wen Chen, Phys. Rev. C **82**, 054607 (2010).

[8] Lie-Wen Chen, Bao-Jun Cai, Che Ming Ko, Bao-An Li, Chun Shen, and Jun Xu, Phys. Rev. C **80**, 014322 (2009).

[9] A. Klimkiewicz *et al.*, Phys. Rev. C **76**, 051603(R) (2007).

[10] A. Carbone, G. Colò,, A. Bracco, L.-G. Cao, P. F. Bortignon, F. Camera, and O. Wieland, Phys. Rev. C **81**, 041301 (2010).

[11] T. Aumann, C. A. Bertulani, F. Schindler, and S. Typel, Phys. Rev. Lett. **119**, 262501 (2017).

[12] W. M. Seif, Phys. Rev. C **74**, 034302 (2006).

[13] M. Ismail, W. M. Seif, A. Adel, and A. Abdurrahman, Nucl. Phys. A **958**, 202 (2017).

[14] Kirandeep Sandhu, Manoj K. Sharma, Amandeep Kaur, and Raj K. Gupta, Phys. Rev. C **90**, 034610 (2014).

[15] W. M. Seif, Phys. Rev. C **91**, 014322 (2015); J. Phys. G: Nucl. Part. Phys. **40**,105102 (2013).

[16] A. N. Andreyev *et al*., Phys. Rev. C**87**, 054311 (2013).

[17] W. M. Seif, M. M. Botros, and A. I. Refaie, Phys. Rev. C **92**, 044302 (2015).

[18] W. M. Seif, M. Ismail, and E. T. Zeini, J. Phys. G: Nucl. Part. Phys. **44**, 055102 (2017).

[19] S. Peltonen, D. S. Delion, and J. Suhonen, Phys. Rev. C **75**, 054301 (2007).

[20] Chang Xu, Zhongzhou Ren, and Yanqing Guo, Phys. Rev. C **78**, 044329 (2008).

[21] W. M. Seif, M. Ismail, A. I. Refaie and Laila H. Amer, J. Phys. G: Nucl. Part. Phys. **43**, 075101 (2016).

[22] Dongdong Ni and Zhongzhou Ren, Phys. Rev. C **93**, 054318 (2016).

[23] M Ismail and A Adel, J. Phys. G: Nucl. Part. Phys. **44**, 125106 (2017).

[24] W. M. Seif and A. Abdurrahman, Chin. Phys. C **42**, 014106 (2018).

[25] Chang Xu, Zhongzhou Ren, and Jian Liu, Phys. Rev. C **90**, 064310 (2014).

[26] Niu Wan, Chang Xu, Zhongzhou Ren, and Jie Liu, Phys. Rev. C **96**, 044331 (2017).

[27] W. M. Seif, N. V. Antonenko, G. G. Adamian, and Hisham Anwer, Phys. Rev. C **96**, 054328 (2017).

[28] D. Vautherin and D. M. Brink, Phys. Rev. C **5**, 626 (1972).

[29] F. L. Stancu and D. M. Brink, Nucl. Phys. A **270**, 236 (1976).

[30] P. Bonche, H. Flocard, and P. H. Heenen, Nucl. Phys. A **467**, 115 (1987).

[31] E. Chabanat, E. Bonche, E. Haensel, J. Meyer, and R. Schaeffer, Nucl. Phys. A 635, 231 (1998).
[32] M. Dutra, O. Lourenço, J. S. Sá Martins, A. Delfino, J. R. Stone, and P. D. Stevenson, Phys. Rev. C **85**, 035201 (2012).
[33] W. M. Seif, Eur. Phys. J. A 38, 85 (2008).
[34] V.Yu. Denisov and W. Noerenberg, Eur. Phys. J. A **15**, 375 (2002).
[35] P.-G. Reinhard, Computational Nuclear Physics, Vol. 1, edited by K. Langanke, J. A. Maruhn, and S. E. Koonin (Springer-Verlag, Berlin, 1990) p. 28.
[36] W. M. Seif and Hesham Mansour, Int. J. Mod. Phys. E **24**, 1550083 (2015).
[37] M. Ismail, W.M. Seif, and H. El-Gebaly, Phys. Lett. B **563**, 53 (2003).
[38] M. Ismail and W. M. Seif, Phys. Rev. C **81**, 034607 (2010).
[39] S. S. Malik and R. K. Gupta, Phys. Rev. C **39**, 1992 (1989).
[40] M. Iriondo, D. Jerrestam, and R. J. Liotta, Nucl. Phys. A **454**, 252 (1986).
[41] R. G. Lovas, R. J. Liotta, A. Insolia, K. Varga, and D. S. Delion, Phys. Rep. **294**, 265 (1998).
[42] I. Tonozuka and A. Arima, Nucl. Phys. A **323**, 45 (1979).
[43] S. M. S. Ahmed, R. Yahaya, and S. Radiman, Rom. Rep. Phys. **65**, 1281 (2013).
[44] M. Onsi, H. Przysiezniak and J. M. Pearson, Phys. Rev. C **50**, 460 (1994).
[45] J. M. Pearson, Y. Aboussir, A. K. Dutta, R. C. Nayak, M. Farine, and F. Tondeur, Nucl. Phys. A **528**, 1 (1991).
[46] J. M. G. Gomez and M. Casas, Few Body Systems, Suppl. **8**, 374 (1995).
[47] J. Friedrich and P. -G. Reinhard, Phys.Rev. C **33**, 335 (1986).
[48] M. Rayet, M. Arnould, F. Tondeur, and G. Paulus, Astron. Astrophys. **116**, 183 (1982).
[49] J. M. Pearson and R. C. Nayak, Nucl. Phys. A **668**, 163 (2000).
[50] F. Tondeur, M. Brack, M. Farine, and J. M. Pearson, Nucl. Phys. **A 420**, 297 (1984).
[51] S. Shlomo, Phys. Atom. Nucl. **73**, 1390 (2010).
[52] B. K. Agrawal, S. Shlomo, and V. K. Au, Phys. Rev. C **72**, 014310 (2005).
[53] P. -G. Reinhard and H. Flocard, Nucl. Phys. A **584**, 467 (1995).
[54] B. A. Brown, G. Shen, G. C. Hillhouse, J. Meng, and A. Trzcińska, Phys. Rev. C **76**, 034305 (2007).
[55] B. A. Brown (unpublished).
[56] B. K. Agrawal, S. Shlomo, and V. Kim Au, Phys. Rev. C **68**, 031304 (2003).
[57] Q. B. Shen, Y. L. Han, and H. R. Guo, Phys. Rev. C **80**, 024604 (2009).
[58] J. Margueron, J. Navarro, and N. Van Giai, Phys. Rev. C **66**, 014303 (2002).
[59] E. Chabanat, E. Bonche, E. Haensel, J. Meyer, and R. Schaeffer, Nucl. Phys. A **627**, 710 (1997).
[60] E. Chabanat, Ph.D. Thesis, University of Lyon, 1995.
[61] J. Meyer, Lectures at the 11th Joliot-Curie School of Nuclear Physics, Maubuisson, September 1993.

[62] T. Lesinski, M. Bender, K. Bennaceur, T. Duguet, and J. Meyer, Phys. Rev. C **76**, 014312 (2007).
[63] P. Klüpfel, P. -G. Reinhard, T. J. Bürvenich, and J. A. Maruhn, Phys. Rev. C **79**, 034310 (2009).
[64] A. W. Steiner, M. Prakash, J. M. Lattimer, and P. J. Ellis, Phys. Rep. **411**, 325 (2005).
[65] P.-G. Reinhard, D. J. Dean, W. Nazarewicz, J. Dobaczewski, J. A. Maruhn, and M. R. Strayer, Phys. Rev. C **60**, 014316 (1999).
[66] S. Köhler, Nucl. Phys. A **258**, 301 (1976).
[67] P. -G. Reinhard (unpublished).
[68] B. Alex Brown, Phys. Rev. Lett. **85**, 5296 (2000).
[69] X. Roca-Maza, M. Centelles, X. Viñas, and M. Warda, Phys. Rev. Lett. **106**, 252501 (2011).
[70] Niu Wan, Chang Xu, and Zhongzhou Ren, Phys. Rev. C **94**, 044322 (2016).
[71] Meng Wang, G. Audi, F. G. Kondev, W. J. Huang, S. Naimi, and Xing Xu, Chin. Phys. C **41**, 030003 (2017).
[72] G. Audi, F. G. Kondev, Meng Wang, W. J. Huang, and S. Naimi, Chin. Phys. C **41**, 030001 (2017).
[73] Daming Deng and Zhongzhou Ren, Phys. Rev. C **96**, 064306 (2017).
[74] Chang Xu, G. Röpke, P. Schuck, Zhongzhou Ren, Y. Funaki, H. Horiuchi, A. Tohsaki, T. Yamada, and Bo Zhou, Phys. Rev. C **95**, 061306(R) (2017).
[75] W. M. Seif, M. Shalaby, and M. F. Alrakshy, *Phys. Rev.* C **84**, 064608 (2011).